\documentclass[preprint,showpacs,preprintnumbers,amsmath,amssymb]{revtex4}

\usepackage{graphicx}
\usepackage{epsfig}
\usepackage{bm}
\usepackage{amsfonts}
\usepackage{subfigure}

\begin{document}
	
\title{Generalized Brans-Dicke inflation with a quartic potential}
		
\author{Behzad Tahmasebzadeh$^{1}$\footnote{behzadtahmaseb@gmail.com} and Kayoomars Karami$^{2}$\footnote{kkarami@uok.ac.ir}}
	
\address{$^1$Department of Physics, Institute for Advanced Studies in Basic Sciences (IASBS), P.O. Box 45195-1159, Zanjan, Iran\\
$^2$Department of Physics, University of Kurdistan, Pasdaran
Street, P.O. Box 66177-15175, Sanandaj, Iran}

    	\begin{abstract}
			Within the framework of Brans-Dicke gravity, we investigate inflation with the quartic potential, $\lambda\varphi^4/4$, in the presence of generalized Brans-Dicke parameter $\omega_{\rm GBD}(\varphi)$. We obtain the inflationary observables containing the scalar spectral index, the tensor-to-scalar ratio, the running of the scalar spectral index and the equilateral non-Gaussianity parameter in terms of general form of the potential $U(\varphi)$ and $\omega_{\rm GBD}(\varphi)$. For the quartic potential, our results show that the predictions of the model are in well agreement with the Planck 2015 data for the generalized Brans-Dicke parameters
$\omega_{\rm GBD}(\varphi)=\omega_0\varphi^{n}$ and $\omega_0e^{b\varphi}$. This is in contrast with both the Einstein and standard Brans-Dicke gravity, in which the result of quartic potential is disfavored by the Planck data.
\end{abstract}
		


\maketitle

\section{Introduction}\label{sec1}

Hot Big Bang theory has outstanding successes in cosmology, for instance describing the cosmic microwave background (CMB) radiation and the light nucleosynthesis. In spit of these successes, it suffers from several problems such as the flatness problem, the horizon problem and also the magnetic mono-pole problem. Inflation theory was suggested to solve all of these problems,
    with the idea that a short period of rapid accelerated
expansion has occurred before the radiation dominated era \cite{Starobinsky:1980te,Sato:1981ds,Sato:1980yn,Guth:1980zm,Linde:1981mu,Albrecht:1982wi}. In addition to solving the problems of the Hot Big Bang cosmology, inflation can provide a rational explanation for the anisotropy observed in the CMB radiation and also in the large-scale structure (LSS) of the
	universe \cite{Mukhanov:1981xt,Hawking:1982cz,Starobinsky:1982ee,Guth:1982ec}.
	In the standard inflationary scenario, a canonical scalar field with self interacting potential $V(\varphi)$ is minimally coupled to the Einstein gravity. Within the framework of standard inflation, viability of different inflationary potentials in light of the observational data has been extensively investigated in the literature (e.g. \cite{Martin1,Martin2,Huang3}). Among other approaches related with a variety of inflationary models, a very promising
approach to inflation is related with the modified theories of gravity known as Brans-Dicke (BD) gravity, in
which a non-canonical scalar field is non-minimally coupled to the gravitational part of the action. Historically, in 1961, Brans and Dicke \cite{Brans:1961sx} introduced a formalism for gravity according to Mach's principle, in which the metric field is coupled to the scalar field to describe the gravitational force.
	BD gravity has been noteworthy frequently since for several reasons. First, a gravitational scalar field appears in BD theory together with the metric tensor, and a fundamental scalar coupled to gravity is an inescapable feature of superstring, supergravity, and M-theories \cite{Callan:1985ia,Lovelace:1986kr,Green1987}. As far as we know, at the experimental point of view the Higgs boson is only the elementary scalar field of the Standard Model. But at the level of theoretical models, in addition to the Higgs field, some other scalar fields also appear in particle physics and in cosmology, such as the superpartner of spin $ 1/2 $ particles in supergravity, the string dilaton appearing in the supermultiplet of the higher-dimensional graviton, or non-fundamental fields like composite bosons and fermion condensates.
	Second, the most potent motivation for the study of BD gravity comes from this reality that the low energy limit of the bosonic string theory equivalent to a BD theory with $ \omega_{\rm BD}=-1 $, also $ \omega_{\rm BD}=-3 $ is  obtained from a less conventional string theory \cite{Callan:1985ia,Lovelace:1986kr}.
	A further interest in BD gravity emanates from the extended and hyperextended inflationary scenarios of the early universe \cite{Faraoni2004}.
	
All mentioned in above motivate us to investigate the cosmic inflation of the early universe within the framework of the BD gravity in which the constant BD parameter is generalized to a function of scalar field, i.e. $\omega_{\rm GBD}(\varphi)$. Our main aim is to examine the viability of the quartic potential $U(\varphi)=\lambda\varphi^4/4$ in light of the Planck 2015
 results. Note that the result of this potential in both the Einstein \cite{Planck2015} and standard BD gravity \cite{Tahmasebzadeh:2016irh} is disfavored by the Planck data. The paper is organized as follows. In section \ref{sec2}, we investigate inflation in the generalized BD setting. We introduce the background equations as well as the scalar and tensor power spectrum. Then we obtain the inflationary observables in terms of the slow roll parameters. In section \ref{sec3}, for a quartic potential with two choices of the generalized BD parameter containing the power-law and exponential functions of the scalar field, we examine the predictions of the model in light of the Planck 2015 data. Section \ref{sec4} is devoted to our conclusions.

\section{Inflation in the generalized BD gravity}\label{sec2}

The action of generalized BD gravity in the Jordan frame is given by \cite{Faraoni2004, Fujii2004, Felice2010, DeFelice:2010jn, Karami:2014tsa,Saridakis}
\begin{equation}
		S =\frac{1}{2} \int d^4 x\sqrt{-g} \left[ \varphi R - \frac{\omega_{\rm GBD} (\varphi )}{\varphi}{g^{\mu \nu }}\partial_\mu\varphi\partial_\nu\varphi-2U(\varphi)\right],
		\label{action}
	\end{equation}
where $R$, $\varphi$, $\omega_{\rm GBD} (\varphi )$ and $U(\varphi)$ are the Ricci scalar, the scalar field, the generalized BD parameter and the self interacting potential, respectively. Note that for the case of $\omega_{\rm GBD}(\varphi)=\omega_0=$ cte., the action (\ref{action}) reduces to the standard BD gravity. Here we take $M_P=(8\pi G)^{-1/2}=1$. For a spatially flat Friedmann-Robertson-Walker (FRW) universe, the Friedmann equations in generalized BD gravity take the forms \cite{Felice2010}.
\begin{align}
	\label{feq1}
		3\varphi H^{2} +3H \dot{\varphi}-\frac{1}{2}\omega(\varphi)\dot{\varphi}^{2}-U(\varphi) &=0 ,\\
		\label{freq2}
			-2 \varphi \dot{H} -\ddot{\varphi}+H\dot{\varphi}-\omega(\varphi) \dot{\varphi}^{2} &=0,
\end{align}
where $\omega(\varphi)\equiv\omega_{\rm GBD}(\varphi)/\varphi$ and the dot denotes a derivative with respect to the cosmic time $t$. Also the continuity equation reads
\begin{align}
\label{coeq}
			\ddot{\varphi}+3 H \dot{\varphi}+\frac{\omega_{,\varphi}}{2 \omega(\varphi)}\dot{\varphi}^{2}+\frac{U_{,\varphi}}{\omega(\varphi)}-\frac{6H^{2}}{\omega(\varphi)}-\frac{3\dot{H}}{\omega(\varphi)}&=0,
\end{align}
where $\omega_{,\varphi}\equiv d\omega/d\varphi$ and $U_{,\varphi}\equiv dU/d\varphi$. Using the slow-roll conditions $ |\dot{\varphi}|\ll |H\varphi| $ and $ |\ddot{\varphi}|\ll |3H\dot{\varphi}|$, Eqs. (\ref{feq1}) and (\ref{coeq}) reduce to
\begin{align}
	\label{Heq}
		3\varphi H^{2} &\simeq U(\varphi),\\
    \label{Hdot}
		3 H \dot{\varphi}&\simeq -\left(\frac{2}{2\omega(\varphi) \varphi+3}\right)\left[ \varphi U_{\varphi}-2U(\varphi)+\omega_{,\varphi}\varphi \dot{\varphi}^{2} \right].
\end{align}
	Replacing $H$ from Eq. (\ref{Heq}) into (\ref{Hdot}) gives
	\begin{equation}
		\dot{\varphi}\simeq\frac{-\sqrt{\frac{3U(\varphi)}{\varphi}}\big(2\omega(\varphi)\varphi+3\big)\pm\sqrt{\frac{3U\big(2\omega(\varphi)\varphi+3\big)^{2}-16\varphi^{2}\big(\varphi U_{,\varphi}-2U(\varphi)\big)\omega_{,\varphi}}{\varphi}}}{4 \varphi~\omega_{,\varphi}},
		\label{scalareq}
	\end{equation}
where we choose the positive sign in Eq. (\ref{scalareq}). Because, our numerical results presented in section \ref{sec3} shows that the negative sign has no end for inflation. Now we turn to calculate the inflationary observable parameters. To this aim, we need to obtain the scalar and tensor power spectrum. Using the perturbed equations in the scalar-tensor gravity which is a general theory that includes the BD gravity, the power spectrum of the curvature perturbation in the slow-roll approximation takes the form \cite{Felice2010}
	\begin{equation}\label{PsH}‎
		‎{{\cal P}_{s}} \simeq \frac{1}{{{Q_s}}}{\left( {\frac{H}{{2\pi }}} \right)^2}\Big|_{k=aH},
		‎\end{equation}
where ‎${\cal P}_{s}$ should be evaluated at the time of horizon exist, i.e. $k=aH$. Here
\begin{equation}\label{Qs}
		{Q_s} \equiv \frac{{\omega (\varphi ){{\dot \varphi }^2} + \frac{{3{{\dot \varphi}^2}}}{{2\varphi}}}}{{{{\left( {H + \frac{{\dot\varphi}}{{2\varphi}}} \right)}^2}}}.
	\end{equation} 	
	The recent value of the scalar perturbation amplitude has been estimated as $ \mathcal{P}_{s} =(2.139 \pm 0.063) \times 10^{-9} $ (Planck 2015 TT,TE,EE+lowP data) \cite{Planck2015}.

	The scale-dependence of the scalar power spectrum is determined by the scalar spectral index $n_{s}$. In the slow roll approximation, it reads
	\begin{equation}‎\label{nsnuR}
		{n_s} - 1 \equiv \frac{{d\ln {{\cal P}_{s}}}}{{d\ln k}}\simeq- 4{\varepsilon _1} - 2{\varepsilon _2} + 2{\varepsilon _3} - 2{\varepsilon _4},
		‎\end{equation}
where $\varepsilon _i,\, (i = 1,2,3,4)$ are the slow-roll parameters defined as \cite{Felice2010,Hwang:2001pu}
	\begin{equation}\label{eps}
		{\varepsilon _1} \equiv  - \frac{{\dot H}}{H^2}, ‎\hspace{.3cm} {\varepsilon _2} \equiv \frac{{\ddot \varphi }}{{H\dot \varphi }}‎, ‎\hspace{.3cm} {\varepsilon _3} \equiv \frac{{\dot \varphi}}{{2H\varphi}}, ‎\hspace{.3cm} {\varepsilon _4} \equiv \frac{{\dot E}}{{2HE}},‎
	\end{equation}
and
	\begin{equation}\label{E}
		E \equiv \omega(\varphi) \varphi+\frac{3}{2}.
	\end{equation}	
The scalar spectral index measured by the Planck 2015 is about $ n_{s}=0.9645 \pm 0.0049 $ (68\% CL) \cite{Planck2015}.

From Eq. (\ref{nsnuR}), one can calculate the running of the scalar spectral index as
	\begin{equation}\label{dnsBD}
		\frac{{d{n_s}}}{{d\ln k}} \simeq  - 8\varepsilon _1^2 + 2\varepsilon _2^2 - 4\varepsilon _3^2 +4\varepsilon _4^{2}- 2{\varepsilon _1}{\varepsilon _2} + 4{\varepsilon _1}{\varepsilon _3}-4{\varepsilon _1}{\varepsilon _4}.
 	\end{equation}
The recent measured value of this parameter is $ d n_{s}/ d \ln k=−0.0057 \pm 0.0071 $ (68\% CL, Planck 2015 TT,TE,EE+lowP data) \cite{Planck2015}.

 The power spectrum of tensor perturbations can be realized in a similar approach that was followed for deriving the scalar perturbations. In the slow-roll regime, it is given by \cite{Felice2010}
	\begin{equation}‎\label{Pt}
		{{\cal P}_t} \simeq \frac{2}{{{\pi ^2}}}\frac{{{H^2}}}{\varphi}\Big|_{k=aH}‎.
		‎\end{equation}
		The tensor spectral index $n_t$ which shows the deviation of the tensor power spectrum from the scale invariance regime, can be obtained as
	\begin{equation}\label{ntst}
		{n_t} \equiv \frac{{d\ln {{\cal P}_t}}}{{d\ln k}}\simeq  - 2{\varepsilon _1} - 2{\varepsilon _3}.
	\end{equation}
Using Eqs. (\ref{PsH}) and (\ref{Pt}), the tensor-to-scalar ratio $r$ in the slow-roll approximation turns into
	\begin{equation}‎\label{rst}
		‎r \equiv \frac{{\cal P}_t}{{\cal P}_{s}}\simeq 8\frac{{{Q_s}}}{\varphi}‎.
		‎\end{equation}
	The recent constraint on this observable has been obtained by Planck satellite as $ r<0.1 $ (95\% CL, Planck 2015 TT,TE,EE+lowP data) \cite{Planck2015}.

Note that although calculations in the Einstein frame is more intuitive, we obtained the inflationary observables in the Jordan frame which is our physical frame. The equivalence between the Einstein frame and the Jordan frame has already been shown for the scalar-tensor theories in \cite{Felice2010,DeFelice:2011jm}. It was pointed out that this equivalence is a consequence of the fact
that both the scalar and tensor spectra, i.e. ${\cal P}_s$ and ${\cal P}_t$, are unchanged under the conformal transformation \cite{Felice2010,DeFelice:2011jm}.
    	
Another important observable predicted by inflation is non-Gaussianity parameter $f_{\rm NL}$ which determines the variance of perturbations from the Gaussian distribution (for review see e.g. \cite{Bartolo:2004if,Chen:2010xka}).
	Different inflationary models predict maximal signal for different shapes of non-Gaussianity. The squeezed shape is the predominant mode of models with multiple light fields during inflation. Also, for the single field inflationary models with non-canonical kinetic terms, the non-Gaussianity parameter has peak in the equilateral shape. Furthermore, the folded non-Gaussianity becomes predominant in models with non-standard initial conditions \cite{Babich:2004gb, Baumann2009}. The equilateral non-Gaussianity parameter
	for the scalar-tensor gravity has been obtained in \cite{DeFelice:2011jm} as
	\begin{equation}
	f_{{\rm{NL}}}^{{\rm{equil}}} \simeq\frac{55}{36}\varepsilon_{s}+\frac{5}{12}\eta_{s},
		\label{fnl}
	\end{equation}
	where
	\begin{equation}
		\varepsilon_{s}\equiv -\frac{\dot H}{H^{2}}+\frac{\dot\varphi}{H\varphi}, \hspace{.5cm} \eta_{s}\equiv \frac{\dot{\varepsilon_{s}}}{H \varepsilon_{s}}.
		\label{fnlvar}
	\end{equation}
	The number of $e$-folds before inflation ends is defined as
\begin{equation}
		N = - \int^{\varphi}_{\varphi_{e}}\frac{H}{{\dot \varphi }}d\varphi,
		\label{efold2}
	\end{equation}
	where $H$ and $\dot{\varphi}$ are given by Eqs. (\ref{Heq}) and (\ref{scalareq}), respectively. Here $\varphi_{e}$ is the scalar field at the end of inflation and it is determined by the condition $\varepsilon_{1}=1$.
	The anisotropies observed in the CMB is equivalent to the perturbations whose wavelengths crossed the Hubble radius around $ N_{*} \approx 50 - 60 $ before the end of inflation \cite{Liddle:2003as,Dodelson:2003vq}.
	In what follows, for the quartic potential $U(\varphi)=\lambda\varphi^4/4$ and two special choices of the generalized BD parameter $\omega_{\rm GBD}(\varphi)$, we obtain the inflationary observables in term of $ \varphi$. Then using the $e$-fold number (\ref{efold2}), we calculate the scalar field $\varphi_{*}$ at the time of horizon exit ($N_{*}=50-60 $), numerically. Therefore, we can plot the $ r-n_{s}$ diagram for the model and examine its viability in light of the Planck 2015 results. In addition, we estimate the running of the scalar spectral index $d{n_s}/d\ln k$ and the equilateral non-Gaussianity $f_{\rm NL}^{\rm equil}$ for our model and compare their results with the observations.

\section{Quartic potential and generalized BD parameter}\label{sec3}

	Here, we consider a quartic potential as follows
		\begin{equation}
		U(\varphi)=\frac{\lambda}{4}\varphi^{4},
		\label{potential}
	\end{equation}
	which is one of the simplest chaotic inflationary potentials \cite{Roberts:1994ap,Racioppia}. Quartic potential in the standard model of inflation which is based on the Einstein gravity, runs into trouble with the CMB \cite{Planck2015}. Because, its prediction for the tensor-to-scalar ratio $r$ is too large and it is in disagreement with the current constraint $r<0.1$ deduced from the Planck 2015 data. Also, the prediction of quartic potential in the standard BD gravity is disfavored by the Planck 2015 results \cite{Tahmasebzadeh:2016irh}.
	 In \cite{Sen:2000zk}, it was shown that in the context of standard BD theory, the potential (\ref{potential}) can justify the late-time accelerated phase of the universe. It was pointed out that the scalar field $ \varphi $ in BD gravity can play the role of the dynamical $\Lambda $ and describe the missing energy. Authors of Ref.\cite{Sen:2000zk} also computed different parameters like the age of the universe, the luminosity-distance redshift relation and the time variation of gravitational coupling and show that the aforementioned cosmological parameters agree quite well with the observations.
	
	It is well known that within the framework of standard BD gravity, the constant BD parameter has constraint $ \omega_{\rm BD}>10^{5} $ from the solar system test. On the other hand, it was shown that the smaller values of this parameter are needed to justify the late time accelerated expansion of the universe driven by dark energy \cite{Li:2015aug}. In \cite{Farajollahi:2011xb}, it was elaborated that in the framework of BD gravity with the scalar field dependent BD parameter $\omega_{\rm GBD}(\varphi)=\omega_{0}\varphi^{n}$, one can unify a decelerating
radiation dominated era in the early time and an accelerated
dark energy dominated era in the late time.
	The generalized BD theory containing a time-dependent BD parameter $\omega_{\rm GBD}(t)$ was introduced by \cite{Nordtvedt:1970uv,Wagoner:1970vr}. The BD theory with time varying
$\omega(t)$ emerges naturally in Kaluza-Klein theories, supergravity theory and in all the well-known effective
	string actions \cite{Green1987,Freund:1982pg}. For some special functions of $\omega(\varphi)$, the generalized BD gravity acts like as graviton-dilaton theory \cite{Russo:1992yg}. In addition, a few attempts have been done to
	study the dynamics of the universe in generalized BD scenario. For instance, for large $ \omega(t)$, BD theory gives the correct amount of inflation and early and late time behavior, and for small negative $ \omega(t) $, it correctly explains cosmic acceleration, structure formation and coincidence problem \cite{Sahoo:2002rx}. In what follows, we consider two choices for the generalized BD parameter as $ \omega_{\rm GBD}(\varphi)=\omega_{0}
	\varphi^{n}$ and $ \omega_{0}e^{b \varphi}$, and examine the viability of the quartic inflationary potential (\ref{potential}) in light of the Planck 2015 results.

\subsection{ Power-law generalized BD parameter}\label{sec3.1}

\begin{figure}
			\centering{	\includegraphics[width=7cm]{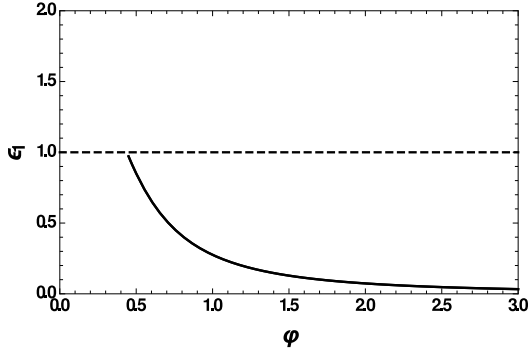}}
			\caption{First slow-roll parameter of quartic potential (\ref{potential}) for the model (\ref{model1}) versus the scalar field. Here $ n=3 $ and $ \omega_{0}=10$. In general, the end of inflation constraint ($\varepsilon_1=1$) is respected for $1< n<4 $ and $ \omega_{0}>0 $.}
			\label{slowroll}
		\end{figure}

\begin{figure}
		\centering{	\includegraphics[width=7cm]{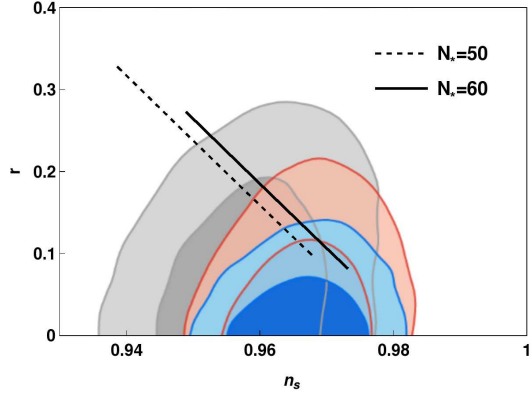}}
		\caption{The result of the quartic potential (\ref{potential}) for the model (\ref{model1}) in $r-n_s$ plane. The marginalized joint 68\% and 95\% CL regions of Planck 2013, Planck 2015 TT+lowP and Planck 2015 TT,TE,EE+lowP data \cite{Planck2015} are specified by gray, red and blue, respectively. The results for $ N_{ *}= 50 $ and $ N_{ *}= 60 $ are shown by the dashed and solid lines, respectively. Here $ 1< n< 4 $ and $\omega_{0}>0$.}
		\label{n-r}
	\end{figure}

\begin{figure}
		\centering{	\includegraphics[width=7cm]{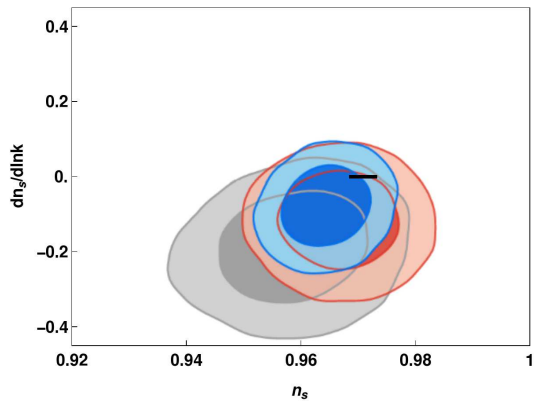}}
		\caption{Prediction of the quartic potential (\ref{potential}) for the model (\ref{model1}) in $ dns/d\ln k-ns $ plane. The grey, red and blue marginalized joint 68\% and 95\% CL regions
			correspond to Planck 2013, Planck 2015 TT+lowP and Planck 2015 TT,TE,EE+lowP data \cite{Planck2015},
			respectively. Here $ 2\leq n< 4 $, $\omega_{0}>0$ and $ N_{*}=60$.}
		\label{running}
	\end{figure}

For the first model of $\omega_{\rm GBD}(\varphi)$, we consider a power-law generalized BD parameter given by \cite{Farajollahi:2011xb,Barrow:1995fj}
	\begin{equation}
		\omega_{\rm GBD}(\varphi)=\omega_{0}\varphi^{n},
		\label{model1}
	\end{equation}
	where $\omega_0$ and $n$ are constant. For the case of $n=0$, Eq. (\ref{model1}) recovers the standard BD gravity. To constrain the parametric space of the model containing $\omega_{0}$ and $n$, we initially check the first slow-roll parameter to satisfy both the slow roll approximation ($\varepsilon_1\ll 1$) during inflation and the condition of end of inflation ($\varepsilon_1=1$). Our numerical results show that inflation ends just for $1<n<4 $ and $\omega_{0}>0 $, otherwise $\varepsilon_{1}$ never arrives to unity. Variations of the first slow-roll parameter $\varepsilon_1$ versus the scalar field is shown in Fig. \ref{slowroll}. We see that during inflation when $\varphi$ decreases, $\varepsilon_1$ increases and then goes to unity at the end of inflation ($\varepsilon_{1}\simeq 1$). 		
 Now with the help of Eq. (\ref{efold2}), we calculate $\varphi_*$ at the horizon exit $e$-fold numbers $ N_{*}=50 $ and 60, numerically. This enable us to obtain the scalar spectral index $n_s$ and the tensor-to-scalar ratio $r$ from Eqs. (\ref{nsnuR}) and (\ref{rst}), respectively, in terms of $\varphi_*$. Figure \ref{n-r} presents the $ r - n_{s} $ diagram for the model (\ref{model1}) with $ N_{*} = 50  $ and $ N_{*} = 60 $ in comparison with the observational data. The results have been plotted for $ 1< n<4 $ and $\omega_0>0$, according to end of inflation constraint ($\varepsilon_1=1$). Note that our numerical calculations show that the results of $r -n_{s} $ diagram are valid for any given values of $\omega_0$ in the range of $\omega_{0}>0$. Figure \ref{n-r} shows that the result of the model (\ref{model1}) for $2\leq n<4 $ lies inside the region 95\% CL of Planck 2015 TT,TE,EE+lowP data \cite{Planck2015}. This is in contrast with the result of quartic potential in both the Einstein \cite{Planck2015} and standard BD gravity \cite{Tahmasebzadeh:2016irh} in which the prediction of model is ruled out by the Planck data. Using Eq. (\ref{dnsBD}), we evaluate the running of the scalar spectral index $d{n_s}/d\ln k$ in our model. Figure \ref{running} shows the result of $d{n_s}/d\ln k$ for $2\leq n<4$ which is compatible with the Planck 2015 data. Also using Eq. (\ref{fnl}), the equilateral non-Gaussianity $f_{{\rm{NL}}}^{{\rm{equil}}}$ for $ 2\leq n< 4 $ and $N_{*}=60 $ is obtained as $ 0.013 < f_{{\rm{NL}}}^{{\rm{equil}}}\leq 0.019 $ which takes place
inside the 68\% CL region of Planck 2015 TT,TE,EE+lowP data \cite{Planck2015}.

\subsection{Exponential generalized BD parameter}\label{sec3.2}

\begin{figure}
 			\centering{	\includegraphics[width=7cm]{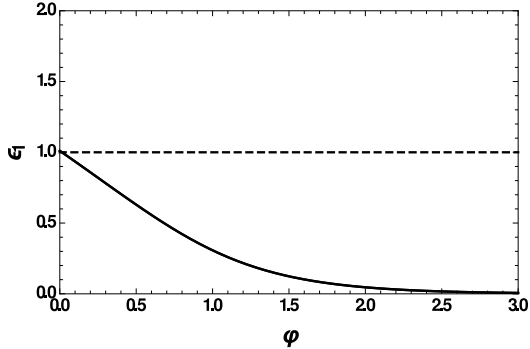}}
 			\caption{Same as Fig. \ref{slowroll}, but for the model (\ref{model2}). Here $b=2$ and $ \omega_{0}=1.2$. In general, the condition $\varepsilon_1=1$ is satisfied for $0.09\leq\omega_{0}\leq 1.2$ and $ b > 0$.}
 			\label{slowroll2}
 		\end{figure}

\begin{figure}
		\centering{	\includegraphics[width=7cm]{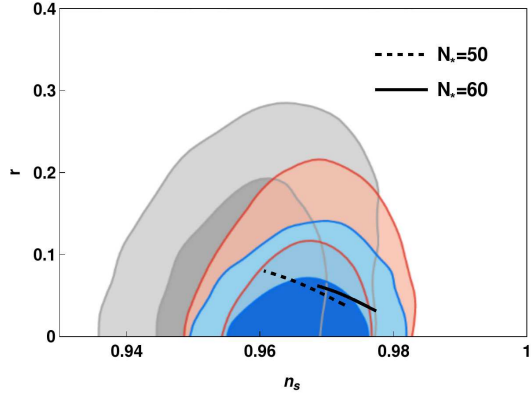}}
		\caption{Same as Fig. \ref{n-r}, but for the model (\ref{model2}). Here $0.09\leq \omega_{0}\leq 1.2$ and $b>0$.}
		\label{1n-r2}
	\end{figure}

\begin{figure}
				\centering{	\includegraphics[width=7cm]{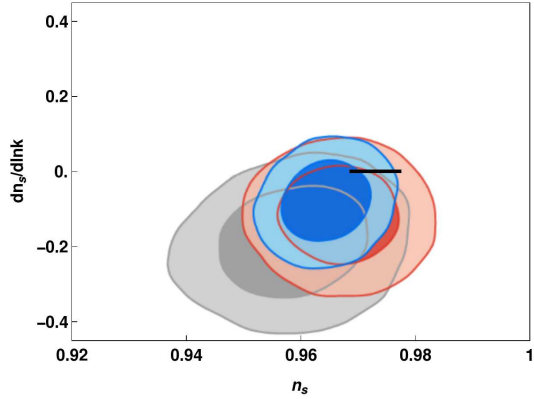}}
				\caption{Same as Fig. \ref{running}, but for the model (\ref{model2}). Here $ 0.09\leq\omega_{0}\leq 1.2$, $b>0$ and $N_{*}=60$.}
				\label{running2}
			\end{figure}

	Secondly, we consider another case of the field-dependent coupling with the kinetic energy as
	\begin{equation}
		\omega_{\rm GBD}(\varphi)=\omega_{0}e^{b\varphi},
		\label{model2}
	\end{equation}
 where $\omega_0$ and $b$ are constant. This exponential coupling term is motivated by the dilatonic coupling in low-energy effective string
 theory \cite{DeFelice:2011jm}. For the case of $b=0$, Eq. (\ref{model2}) turns into the standard BD model. Here, the free parameters $\omega_{0}$ and $b$ are constrained from the end of inflation constraint, i.e. $\varepsilon_1=1$. This limits our parametric space to $ 0.09\leq\omega_{0}\leq 1.2$ and $ b > 0$, see Fig. \ref{slowroll2}.
 The $ r-n_{s} $ diagram for the model (\ref{model2}) with $ N_{*} = 50  $ and $ N_{*} = 60 $ is plotted in Fig. \ref{1n-r2}. Note that the numerical result of $ r-n_{s} $ diagram is independent of $b$. We need just to have $b>0$ due to having the end of inflation. Figure \ref{1n-r2} shows that, in contrary to the result of quartic potential in both the Einstein \cite{Planck2015} and standard BD gravity \cite{Tahmasebzadeh:2016irh}, the result of the model (\ref{model2}) for $ 0.09\leq \omega_{0}\leq 1.2 $ lies inside the 68\% CL region of Planck 2015 TT,TE,EE+lowP data \cite{Planck2015}. Also the running of the scalar spectral index predicted by the model (\ref{model2}) is favored by the Planck 2015 data, see Fig. \ref{running2}. Furthermore, the equilateral non-Gaussianity $f_{{\rm{NL}}}^{{\rm{equil}}}$ for $ 0.09\leq\omega_{0}\leq 1.2 $ is obtained as $ 0.010 \leq f_{{\rm{NL}}}^{{\rm{equil}}}\leq 0.015 $ which lies inside the 68\% CL region of Planck 2015 TT,TE,EE+lowP data \cite{Planck2015}.
		
\section{Conclusions}\label{sec4}

Here, we investigated inflation driven by the quartic potential $U(\varphi)=\lambda\varphi^4/4$ in the framework of generalized BD theory with a scalar field dependent BD parameter $\omega_{\rm GBD}(\varphi)$. First, we obtained the necessary relations for the inflationary observables containing the scalar spectral index $n_s$, the tensor-to-scalar ratio $r$, the running of the scalar spectral index $dn_s/d\ln k$ and the equilateral non-Gaussianity $f_{\rm NL}^{\rm equil}$ in terms of general functions of $U(\varphi)$ and $\omega_{\rm GBD}(\varphi)$. Then, for the quartic potential $U(\varphi)=\lambda\varphi^4/4$ with the two choices of $ \omega_{\rm GBD}(\varphi)=\omega_{0}
	\varphi^{n}$ and $ \omega_{0}e^{b \varphi}$, we examined the viability of the models in light of the Planck 2015 data. Note that the result of the quartic potential in both the Einstein and standard BD gravity ($\omega_{\rm GBD}(\varphi)=\omega_{0}$) is disfavored by the Planck data. For the model $ \omega_{\rm GBD}(\varphi)=\omega_{0}\varphi^{n}$, the result of $r-n_s$ diagram for $2\leq n<4 $ and $\omega_0>0$ lies inside the region 95\% CL of Planck 2015 TT,TE,EE+lowP data \cite{Planck2015}. The result of $r-n{_s}$ for another model $ \omega_{\rm GBD}(\varphi)=\omega_{0}e^{b \varphi}$ with $0.09\leq \omega_{0}\leq 1.2$ and $b>0$ takes place in the 68\% CL region of Planck 2015 TT,TE,EE+lowP data.
For the both $ \omega_{\rm GBD}(\varphi)=\omega_{0}
	\varphi^{n}$ and $ \omega_{0}e^{b \varphi}$, the
prediction of the running of the scalar spectral index $dn_s/d\ln k$ is compatible with the Planck 2015 data. Furthermore, the equilateral non-Gaussianity predicted by the both models lies inside the 68\% CL region of Planck 2015 TT,TE,EE+lowP data \cite{Planck2015}.

\subsection*{Acknowledgements}
 Behzad Tahmasebzadeh would like to thank Kazem Rezazadeh for useful discussions.



\begin{thebibliography}{}

	\expandafter\ifx\csname natexlab\endcsname\relax\def\natexlab#1{#1}\fi
	\expandafter\ifx\csname bibnamefont\endcsname\relax
	\def\bibnamefont#1{#1}\fi
	\expandafter\ifx\csname bibfnamefont\endcsname\relax
	\def\bibfnamefont#1{#1}\fi
	\expandafter\ifx\csname citenamefont\endcsname\relax
	\def\citenamefont#1{#1}\fi
	\expandafter\ifx\csname url\endcsname\relax
	\def\url#1{\texttt{#1}}\fi
	\expandafter\ifx\csname urlprefix\endcsname\relax\def\urlprefix{URL }\fi
	\providecommand{\bibinfo}[2]{#2}
	\providecommand{\eprint}[2][]{\url{#2}}
	
\bibitem{Starobinsky:1980te}
A.A.~Starobinsky,
Phys.\ Lett.\ B {\bf 91}, 99 (1980).

\bibitem{Sato:1981ds}
K.~Sato,
Phys.\ Lett.\ B {\bf 99}, 66 (1981).

\bibitem{Sato:1980yn}
K.~Sato,
Mon.\ Not.\ Roy.\ Astron.\ Soc.\  {\bf 195}, 467 (1981).
	
\bibitem{Guth:1980zm}
A.H.~Guth,
Phys.\ Rev.\ D {\bf 23}, 347 (1981).
	
\bibitem{Linde:1981mu}
A.D.~Linde,
Phys.\ Lett.\ B {\bf 108}, 389 (1982).
	
\bibitem{Albrecht:1982wi}
A.~Albrecht, P.J.~Steinhardt,
Phys.\ Rev.\ Lett.\  {\bf 48}, 1220 (1982).
	

\bibitem{Mukhanov:1981xt}
V.F.~Mukhanov, G.V.~Chibisov,
JETP Lett.\  {\bf 33}, 532 (1981).
	
\bibitem{Hawking:1982cz}
S.W.~Hawking,
Phys.\ Lett.\ B {\bf 115}, 295 (1982).
	
\bibitem{Starobinsky:1982ee}
A.A.~Starobinsky,
Phys.\ Lett.\ B {\bf 117}, 175 (1982).

\bibitem{Guth:1982ec}
A.H.~Guth, S.Y.~Pi,
Phys.\ Rev.\ Lett.\  {\bf 49}, 1110 (1982).
	
\bibitem{Martin1} J. Martin, C. Ringeval, V. Vennin, Phys. Dark Univ. \textbf{5-6}, 75 (2014).

\bibitem{Martin2} J. Martin, C. Ringeval, R. Trotta, V. Vennin, JCAP \textbf{03}, 039 (2014).

\bibitem{Huang3} Q.G. Huang, K. Wang, S. Wang, Phys. Rev. D \textbf{93}, 103516 (2016).

\bibitem{Brans:1961sx}
C.~Brans, R.H.~Dicke,
Phys.\ Rev.\  {\bf 124}, 925 (1961).

\bibitem{Callan:1985ia}
C.G.~Callan, D.~Friedan, E.J.~Martinec, M.J.~Perry,
Nucl.\ Phys.\ B {\bf 262}, 593 (1985).

\bibitem{Lovelace:1986kr}
C.~Lovelace,
Nucl.\ Phys.\ B {\bf 273}, 413 (1986).

\bibitem{Green1987}
B. Green, J.M.~Schwarz, E.~Witten,
\textit{Superstring Theory}, Cambridge University Press, Cambridge (1987).

\bibitem{Faraoni2004}
V.~Faraoni,
\textit{Cosmology in Scalar-Tensor Gravity}, Springer Netherlands, Dordrecht (2004).
	
\bibitem{Planck2015}
P.A.R. Ade et al. (Planck collaboration),
A\&A \textbf{594}, A20 (2016).	

\bibitem{Tahmasebzadeh:2016irh}
B.~Tahmasebzadeh, K.~Rezazadeh, K.~Karami,
JCAP {\bf 07}, 006 (2016).
	
\bibitem{Felice2010}
A. De Felice, S.~Tsujikawa,
Living Rev. Relativ. {\bf 13}, 3 (2010).	
	



	
\bibitem{Fujii2004}
Y.~Fujii, K.~Maeda,
\textit{The Scalar-Tensor Theory of Gravitation}, Cambridge University Press, Cambridge (2004).	
		

\bibitem{DeFelice:2010jn}
A. De Felice, S.~Tsujikawa,
JCAP {\bf 07}, 024 (2010).

\bibitem{Karami:2014tsa}
A.~Abdolmaleki, T.~Najafi, K.~Karami,
Phys.\ Rev.\ D {\bf 89}, 104041 (2014).

\bibitem{Saridakis} G. Kofinas, E. Papantonopoulos, E.N. Saridakis, Class. Quantum Grav. {\bf 33}, 155004 (2016).


\bibitem{Hwang:2001pu} J.C.~Hwang, H.~Noh,
Phys.\ Lett.\ B {\bf 506}, 13 (2001).


\bibitem{DeFelice:2011jm}
A. De Felice, S.~Tsujikawa, J.~Elliston, R.~Tavakol,
JCAP {\bf 08}, 021 (2011).	




\bibitem{Bartolo:2004if}
N.~Bartolo, E.~Komatsu, S.~Matarrese, A.~Riotto,
Phys.\ Rept.\  {\bf 402}, 103 (2004).
		
\bibitem{Chen:2010xka}
X.~Chen,
Adv.\ Astron.\  {\bf 2010}, 638979 (2010).	

\bibitem{Babich:2004gb}
D.~Babich, P.~Creminelli, M.~Zaldarriaga,
JCAP {\bf 08}, 009 (2004)

\bibitem{Baumann2009}
D.~Baumann,
arXiv:0907.5424.	


				
\bibitem{Liddle:2003as}
A.R.~Liddle, S.M.~Leach,
Phys.\ Rev.\ D {\bf 68}, 103503 (2003).
	
\bibitem{Dodelson:2003vq}
S.~Dodelson, L.~Hui,
Phys.\ Rev.\ Lett.\  {\bf 91}, 131301 (2003).
	
\bibitem{Roberts:1994ap}
D.~Roberts, A.R.~Liddle, D.H.~Lyth,
Phys.\ Rev.\ D {\bf 51}, 4122 (1995).	

\bibitem{Racioppia} K. Kannike, A. Racioppi, M. Raidal, JHEP {\bf 01}, 035 (2016).


\bibitem{Sen:2000zk}
S.~Sen, A.A.~Sen,
Phys.\ Rev.\ D {\bf 63}, 124006 (2001).

\bibitem{Li:2015aug}
J.X.~Li, et al.,
Res.\ Astron.\ Astrophys.\  {\bf 15}, 2151 (2015).

\bibitem{Farajollahi:2011xb}
H.~Farajollahi, A.~Salehi, F.~Tayebi,
Can.\ J.\ Phys.\  {\bf 89}, 915 (2011).

\bibitem{Nordtvedt:1970uv}
K.~Nordtvedt,
Astrophys.\ J.\  {\bf 161}, 1059 (1970).

\bibitem{Wagoner:1970vr}
R.V.~Wagoner,
Phys.\ Rev.\ D {\bf 1}, 3209 (1970).

\bibitem{Freund:1982pg}
P.G.O.~Freund,
Nucl.\ Phys.\ B {\bf 209}, 146 (1982).

\bibitem{Russo:1992yg}
J.G.~Russo, A.A.~Tseytlin,
Nucl.\ Phys.\ B {\bf 382}, 259 (1992).

\bibitem{Sahoo:2002rx}
B.K.~Sahoo, L.P.~Singh,
Mod.\ Phys.\ Lett.\ A {\bf 17}, 2409 (2002).

\bibitem{Barrow:1995fj}
J.D.~Barrow,
Phys.\ Rev.\ D {\bf 51}, 2729 (1995).

\end{thebibliography}
\end{document}